\documentclass[epsf,twocolumn,showpacs,preprintnumbers]{revtex4}
\usepackage{graphics}
\usepackage{graphicx}
\usepackage{dcolumn} 
\usepackage{bm}
\usepackage{epsfig}
\begin{document}
\title{Disproportionation Transition at Critical Interaction Strength: 
Na$_{1/2}$CoO$_2$}
\author{K.-W. Lee$^1$, J. Kune\v{s}$^{1,2}$, P. Novak$^2$ and W. E. Pickett$^1$}
\affiliation{$^1$Department of Physics, University of California, Davis, CA 95616, 
USA\\ $^2$Institute of Physics, Academy of Sciences,
   Cukrovarnick\'a 10, CZ-162 53 Prague 6, Czech Republic}
\date{\today}
\begin{abstract}
Charge disproportionation (CD) and spin differentiation
in Na$_{1/2}$CoO$_2$ are studied using
the correlated band theory approach.  The simultaneous CD and gap opening
seen previously is followed through a first order 
charge disproportionation transition 
2Co$^{3.5+} \rightarrow$ Co$^{3+}$+Co$^{4+}$, whose ionic identities are
connected more closely to spin (S=0, S=$\frac{1}{2}$ respectively) 
than to real charge.  Disproportionation in the Co $a_g$ orbital is
compensated by opposing charge rearrangement in other $3d$ orbitals. 
At the transition large and opposing
discontinuities in the (all-electron) kinetic and potential energies are
slightly more than balanced by a gain in correlation energy.  The 
CD state is compared to characteristics of the observed charge-ordered 
insulating phase in Na$_{1/2}$CoO$_2$, suggesting the Coulomb repulsion
value $U$ is concentration-dependent, with $U(x=1/2)\simeq$3.5 eV.
\end{abstract}
\maketitle

The discovery by Takada and coworkers of superconductivity\cite{takada} in 
Na$_{0.3}$CoO$_2 \bullet$1.3 H$_2$O near 5 K has led to extensive studies
of the rich variation of properties in the Na$_x$CoO$_2$ system 
($0.2 \le x \le 1$).  It has been pointed out\cite{ucd1,foo,ucd2} that 
the formal charge of the Co ions that occupy a triangular lattice, whose simple
average is $4-x$, changes at $x\simeq 0.5$: for $x > 0.5$ 
the observed Curie-Weiss (C-W) susceptibility indicates 
($1-x$) Co$^{4+}$ ions with
spin S=$\frac{1}{2}$ and $x$ Co$^{3+}$ with S=0, while for $x < 0.5$ 
the susceptibility is Pauli-like and
all Co ions are identical (``Co$^{(4-x)+}$'').  In a narrow region around
$x = 0.5$ there is  
charge ordering (CO) as observed by electron 
diffraction,\cite{zandbergen,foo,huang1}
probably accompanied by antiferromagnetic spin ordering.
While the rest of the phase
diagram is metallic, this $x=0.5$ phase undergoes a CO+metal-insulator 
transition (MIT) at 50 K as verified by 
resistivity\cite{foo} and optical conductivity.\cite{j.hwang,wang}

The mechanism of charge ordering has been a central question in many classes
of transition metal oxides, especially the cuprates and manganates.  The
CO question is beginning to be addressed in these cobaltates, from the
mechanism of ordering at commensurate concentrations\cite{ucd1,motrunich,ucd2}
to the effects of fluctuations when the concentration varies from a
commensurate fraction.\cite{ucd1,baskaranCO}  Preceding the question of
CO however must come the more basic one of disproportionation of a lattice
of identical Co$^{(3+x)+}$ ions for $x < 0.5$ to disproportionated
Co$^{3+}$/Co$^{4+}$ ions with their distinct charges and spins.  
CD is a signature of intraatomic correlation and is closely associated
with local moment formation,
but it also has a collective nature to it (every ion disproportionates).  Moment
formation has been studied primarily in single band models in conjunction
with the correlation-driven MIT,\cite{MIT} 
but CD in multiband
systems (as Na$_{0.5}$CoO$_2$ almost certainly is) is a substantially
more involved question and is relatively unstudied.

Recently some of the present authors obtained a CD/MIT\cite{ucd2}
in a study of Na$_x$CoO$_2$ ($x=1/3$, also $x=2/3$) using the correlated band 
theory method LDA+U (local density approximation plus Hubbard U).
For no apparent reason, the CD/CO transition appeared 
{\it simultaneously} with the MIT (band gap opening) as 
the interaction strength $U$ was
varied in the 3-4 eV range.
Another feature that was not clarified 
is that this CD involved rather little actual redistribution
of charge but at $x=\frac{1}{3}$ showed up dramatically 
in the magnetic moment as the 
metamorphosis 3Co$^{3\frac{2}{3}+} \rightarrow$ Co$^{3+}$+2Co$^{4+}$.
One of the central questions in these cobaltates is the relevance of
its multiband nature, a complicating aspect that has stimulated 
studies into the possibility of
orbital-selective metal-insulator transitions.\cite{liebsch,koga} 
In Na$_x$CoO$_2$ the threefold $t_{2g}$ manifold is partially filled,
with broken symmetry $t_{2g} \rightarrow a_g + e^{\prime}_g$ arising from
the CoO$_2$ layered structure.  LDA calculations indicate both $a_g$
and $e_g^{\prime}$ Fermi surfaces already at $x=\frac{3}{4}$, and the
presence of $e_g^{\prime}$ Fermi surfaces has become a central component
of several models of the superconducting phase.\cite{sctheory}

Another important question is that of
correlation effects and especially the mechanism of CD/CO/MIT transitions.
There have been a few pioneering applications of correlated band theories 
to model disproportionation.\cite{temmerman,ucd1,ucd2} 
In this paper we study more thoroughly the $x=\frac{1}{2}$ case, which has
become more compelling since a CO/MIT was observed.\cite{foo} 
The variable in our study will
 be $U$, but in a small
parameter range such as near the critical value $U_c$, varying $U$ will be
analogous to applying pressure to change the $U/W$ ratio, hence 
predictions are open
to direct experimental comparison.  In addition, the CD/CO that is 
observed as $x\rightarrow 0.5$ suggest a concentration dependent
interaction strength $U(x)$.  We study in detail a mean-field but
fully selfconsistent treatment as provided by LDA+U, of a first order CD
transition driven by correlation, and show that the multiband nature of 
Na$_x$CoO$_2$
is an integral ingredient of the CD transition.  We compare with data
to suggest the value of $U(x)$.

Our calculations that are based on the underlying hexagonal structure 
(space group $\it{P6_322}$) having
lattice constants ($a=2.84$ $\AA$, $c=10.81$ $\AA$),\cite{jansen}
were done with a two Co supercell 
with space group $\it{P2/m}$) allowing two ions 
Co1 and Co2 to be realized in the cell. 
The O height was taken as $z_0=0.168(c/2)=0.908$ $\AA$ as relaxed by 
LDA calculation by Singh.\cite{singh}
The calculations were carried out with the full-potential nonorthogonal 
local-orbital minimum-basis scheme (FPLO) \cite{klaus}
and both popular schemes\cite{anisimov,czyzyk} for LDA+U functional
with the main results being common to each. Since Na$_x$CoO$_2$ is a good
metal (except at $x=\frac{1}{2}$ which we address), the results 
quoted in this paper
are those from the so-called ``around mean field" scheme \cite{anisimov}
appropriate for relatively small $U$. 
The Brillouin zone was sampled with 128 (306 for LDA+U calculations) 
${\bf k}$ points in the irreducible wedge.
Orbitals for the basis set contained $\it{2s2p3s3p3d}$
for Co, O, and Na provide excellent basis flexibility.


\begin{figure}[tbp]
\rotatebox{-90}{\resizebox{6cm}{7cm}{\includegraphics{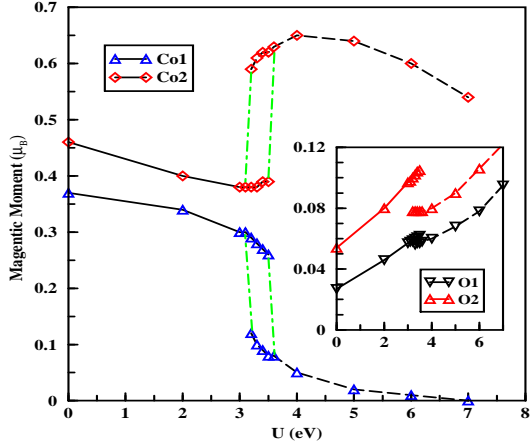}}}
\caption{Effect of $U$ on the magnetic moments for Na$_x$CoO$_2$.
The jump reflects a first-order transition (and accompanying hysteresis)
 in the critical region
(3.2 eV to 3.6 eV).  Gap opening (metal-insulator transition)
simultaneously occurs with CD due to the first-order nature.  
Dashed and solid lines represent the CD and the undisproportionated
state, respectively.
$\it{Insert}$: Change of the magnetic moments for the O ions:
O1 (with 2n site symmetry) is shared by two Co1 and one Co2, while
O2 (with 2m site symmetry) is shared by one Co1 and two Co2.}
\label{MU}
\end{figure}

The doubled unit cell allows for CD into Co$^{3+}$ and Co$^{4+}$,
but with only one magnetic ion there is no issue of spin order 
(except simple alignment).  Hence we work throughout with a ferromagnetic (FM)
doubled cell.  In the local spin density approximation a FM 
state is favored (by a small energy) over the nonmagnetic state,
with a total moment of 1 $\mu_B$ as noted first by Singh.\cite{singh}
Because the Na ions are ordered above the CO transition,\cite{NaCO}
the Na site is nonsymmetric with respect to the Co sites, and the Co ions
have somewhat different moments because the symmetry is already broken
by the Na ordering.

\begin{figure}[tbp]
\rotatebox{-90}{\resizebox{7cm}{7cm}{\includegraphics{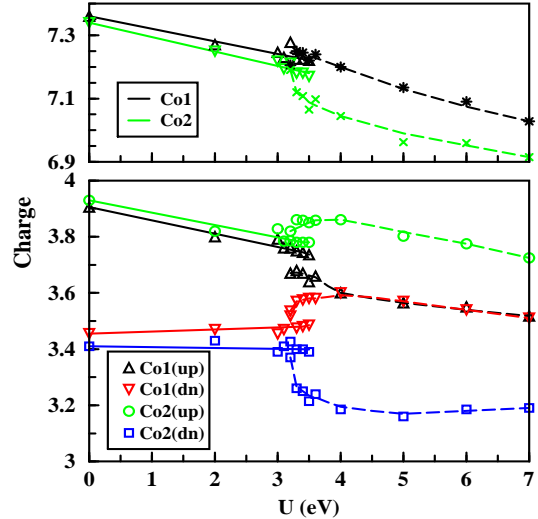}}}
\caption{Change with $U$ of $d$ charge for the two Co ions.
{\it Upper panel}: change of total $3d$ charge (the charge is
transferred to O ions).
{\it Lower panel}: changes of the majority
and minority separately.
}
\label{charge}
\end{figure}

The effect of the on-site Coulomb repulsion $U$ on the
magnetic moments is displayed in Fig. \ref{MU}. 
We follow behavior with $U$ continuously by using the solution from
one value of $U$ as the starting point for the case with incrementally
increased/decreased $U$.  Although unconstrained, the total cell moment
remains at 1 $\mu_B$: half metallic in the metallic
phase, insulating
in the CD phase.
Both Co magnetic 
moments decrease slowly (-0.025 $\mu_B$/eV)
with increasing $U$, probably due to
some charge transfer Co $\rightarrow$ O.
At U=U$_{c2}$=3.6 eV, the solution changes character discontinuously,
with the small and large moments providing the clear identification 
Co1$\rightarrow$Co$^{3+}$, Co2$\rightarrow$ Co$^{4+}$.  By $U$ = 4.5 eV
the large moment obtains a maximum and the small moment is 
negligible.  For larger $U$ the Co$^{4+}$ moment tends to decrease.
Upon decreasing $U$ the moments vary 
continuously down to $U = U_{c1}$ = 3.2 eV, below which the solution
jumps back to the initial undisproportionated state.  Hence we
find a 0.4 eV hysteresis in this CD transition centered at $U_c$
=3.4 eV.  The first order 
nature is consistent with no symmetry having been broken in our
choice of simulation cell, and also accounts for the simultaneous 
CD and gap opening observed previously\cite{ucd2}: the regime of 
accelerating disproportionation includes the opening of the gap, but
this region is inaccessible because of its higher energy than either the
undisproportionated or CD states. 
This first order transition is analogous to low-spin $\leftrightarrow$
high-spin transitions mapped out using 
fixed spin moment calculations,\cite{fixedspin}
but generalizing to a two dimensional space of Co1 and Co2 moments
$\mu_1, \mu_2$ respectively as considered first by Moruzzi, Marcus
and K\"ubler.\cite{moruzzi}  

Although Co1 and Co2 become quite different above $U_c$,
the disproportionation does not show up as strongly in Co$^{4+}
\rightarrow$ Co$^{3+}$ charge transfer
as in the spin rearrangement.
As $U$ increases from zero, a small amount of charge from both Co ions transfers
(0.04-0.05 $e$/eV) onto the O ions (this shift is smaller if using the
``fully localized limit'' LDA+U functional\cite{czyzyk}).  
At the transition, as shown in Fig. \ref{charge}, the  
Co1$\rightarrow$Co$^{3+}$ charge is continuous at
the transition, while the Co2$\rightarrow$Co$^{4+}$ charge 
jumps by $\sim$0.1 electron.  The discontinuities in
the majority and minority charges separately
are also pictured in Fig. \ref{charge}. 

\begin{figure}[tp]
\rotatebox{-90}{\resizebox{7cm}{7cm}{\includegraphics{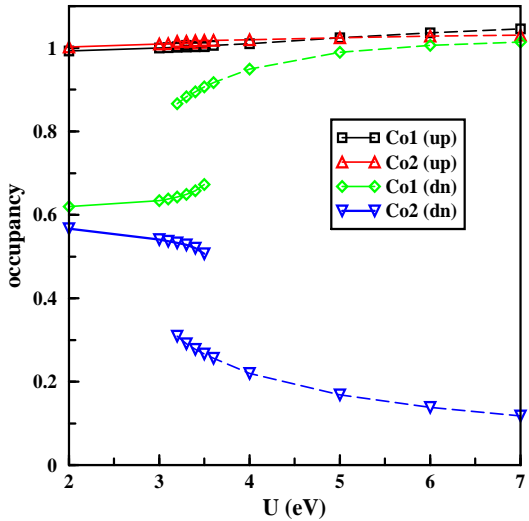}}}
\caption{Change of the occupancy $n_{a_g}$ of the $a_g$ state versus
 $U$, which reveals the strong $a_g$ charge disproportionation
at the transition.  Majority orbitals of both Co ions
 are fully occupied
 regardless of $U$. At $U = U_c$, the occupancy of
 the minority decreases for Co2 and increases for Co1
 by 0.23 $e$.}
\label{occ}
\end{figure}

The picture of the charge 
rearrangement at the transition becomes clearer when the $a_g$ occupation
is monitored, as shown in Fig. \ref{occ}.  At the CD transition, much of
the {\it minority} $a_g$ charge on Co1 ($\rightarrow$ Co$^{3+}$) transfers to
the {\it majority} on Co2 (becoming Co$^{4+}$).  The {\it total charge}
difference is only a fraction of the change in minority $a_g$ charge, because
it is compensated rather strongly by a rehybridization 
and back polarization of the other
$3d$ ($e_g^{\prime}$ and $e_g$) states, reminiscent of what was seen
for variation of the Na content.\cite{marianetti}
A large change in spin is achieved without large change in total
charge because of the multiband nature of the system at and near the
Fermi level.

\begin{figure}[tp]
\rotatebox{-90}{\resizebox{7cm}{7cm}{\includegraphics{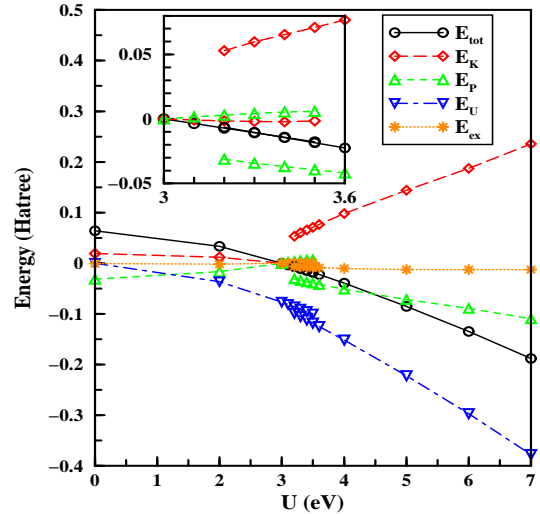}}}
\caption{
 Change in energy E(U) per doubled supercell versus U, relative to
  E(U=3 eV), showing the various contributions
  discussed in the text.
 Values shown are energy differences
 with respect to the value at $U=3$ eV,
 except the LDA+U energy E$_U$ whose actual value is plotted. 
$\it{Insert}$:
 Blowup of the critical region.}
\label{EU}
\end{figure}

The energy E(U)
is shown in Fig. \ref{EU}.
In the critical region, the undisproportionated state is favored
marginally (only 8 meV/Co),
a difference so small that the total energy is essentially continuous
at the transition. 
As CD occurs, the (Kohn-Sham) kinetic energy jumps sharply 
by $0.74$ eV/Co while other contributions decrease:
(in eV/Co) $0.46$ for the (electron+nuclear) potential energy, 
$0.09$ for the LDA exchange, and 0.18 for the 
LDA+U correlation energy ($\propto U$). The LDA+U correction is modest 
but essential
to the transition, and only becomes large enough to tip the balance 
at $U = U_c$.

Now we discuss the relation of our study to Na$_{1/2}$CoO$_2$.  Na
orders in this system above the MIT\cite{foo} 
so our Na-ordered cell is
realistic in that respect.  The observed cell is twice as large as
we have allowed, so it is not possible to study AFM ordering 
in our cell. 
In $x=\frac{1}{2}$ samples, Co$^{3+}$ and Co$^{4+}$ ions are 
disproportionated (C-W susceptibility) but charge/spin
disordered above the transition; conduction is poor
and abruptly becomes insulating when CO occurs.
In our (T=0) simulation, Co1 and Co2 are undisproportionated with a
half metallic spectrum for $U < U_c$, whereas for $U > U_c$ the
ions disproportionate.  Within LDA+U, the MIT is 
orbital-selective.\cite{liebsch,koga}  While the distinction
between our results and cobaltate behavior are evident, 
the question of CD is quite
separate from that of CO: experimentally, the C-W susceptibility indicates
CD in the sample at high temperature, but the
simultaneous CO/MIT occurs only at 50 K.

From the infrared conductivity Hwang {\it et al.} have 
reported\cite{j.hwang} a very small 10 meV gap in their 
insulating $x=\frac{1}{2}$ 
sample.  This small value suggests the value of $U$ required to model
the $x=\frac{1}{2}$ system is very close to (only slightly larger than)
$U_c$, {\it i.e.}
$U(x=\frac{1}{2}) \simeq $3.5-4 eV.  From susceptibility, heat capacity, and 
resistivity data the $x$=0.7-0.8 phases are clearly more correlated
($U >$ 4 eV), confirmed by the very large field-dependent 
thermopower\cite{thermopower}, an extremely large 
Kadowaki-Woods ratio\cite{scattering},
and many reports of magnetic ordering.
On the other hand, in the superconducting phase $x\approx 0.3$ the
resistivity is lower and the mass enhancement over the band value is 
very small\cite{ucd1,ucd2} so this phase is
at most weakly correlated with $U(x\simeq 0.3) <$ 3 eV. 

Whereas previous LDA+U studies\cite{otherLDAU}
of the Na$_x$CoO$_2$ system did not
obtain disproportionation, we have shown that reducing constraints 
allows, and leads to, disproportionation.  
In opposition to LDA+U conventional wisdom
(``precise value of $U$ is not very important''), results 
for Na$_x$CoO$_2$ are very sensitive to the value of $U$ around
the critical value $U_c = 3.4$ eV.
We view Na$_x$CoO$_2$ in terms of a 
concentration dependent $U(x)$, which may be in part a guise for a
single band $\rightarrow$ three band crossover around $x=\frac{1}{2}$
as discussed earlier.\cite{ucd2}  In  such cases studies\cite{OG} of the
multiband Hubbard model suggest $U_{eff}$ for the single band regime
($x$ close to unity) becomes $U_{eff} = U_c/\sqrt{3}$ for the three 
band regime ($x < \frac{1}{2})$, in which case $U$ for the superconducting
phase could be even smaller than our estimate above.

We acknowledge helpful communications with
H. Alloul,
R. J. Cava, A. Eguiluz,
J. D. Jorgensen, M. D. Johannes, R. T. Scalettar,
D. J. Singh, R. R. P. Singh, and
J. M. Tarascon.
This work was supported by DOE grant DE-FG03-01ER45876 and DOE's
Computational Materials Science Network, and by
Czech-USA Project KONTAKT-ME-547.
W. E. P. acknowledges support of the Department of Energy's
Stewardship Science Academic Alliances Program.

\end{document}